\documentclass[aip,jcp,reprint]{revtex4-1}

\usepackage{amsmath,amstext,amssymb,amsopn,amsthm}
\usepackage{color}
\usepackage{graphicx}
\usepackage{enumitem}
\usepackage[normalem]{ulem}

\newcommand{\R}{\mathbb{R}}
\newcommand{\Z}{\mathbb{Z}}

\newcommand{\tf}{\widetilde{f}}

\begin{document}


\title[{Solid Harmonic Scattering Invariants}]{{Solid Harmonic Wavelet Scattering for Predictions of Molecule Properties}}

\author{Michael Eickenberg}
\email{\texttt{michael.eickenberg@nsup.org}}

\author{Georgios Exarchakis}
\email{\texttt{georgios.exarchakis@ens.fr}}

\affiliation{Department of computer science, \'{E}cole
  normale sup\'{e}rieure, PSL Research University, 75005 Paris, France}

\author{Matthew Hirn}
\email{\texttt{mhirn@msu.edu}}

\affiliation{Department of Computational Mathematics, Science \&
  Engineering, Department of Mathematics, Michigan State University,
  East Lansing, Michigan, USA}

\author{St\'{e}phane Mallat}

\affiliation{Coll\`{e}ge de France, \'{E}cole
  normale sup\'{e}rieure, PSL Research University, 75005 Paris, France}
  
\author{Louis Thiry}
  
\affiliation{Department of computer science, \'{E}cole
  normale sup\'{e}rieure, PSL Research University, 75005 Paris, France}

\date{\today}

\begin{abstract}
\textbf{Abstract.}
We present a machine learning algorithm for the prediction of molecule properties inspired by ideas from density functional theory. 
Using Gaussian-type orbital functions, we create surrogate electronic densities of the molecule from which we compute invariant "solid harmonic scattering coefficients" that account for different  types of interactions at different scales. 
Multi-linear regressions of various physical  properties of molecules are 
computed from these invariant coefficients.
Numerical experiments show that these regressions have near state of the art performance, even with relatively few training examples. Predictions over small sets of scattering coefficients can reach a DFT precision while being interpretable.
\end{abstract}

\pacs{02.30.Nw, 02.60.Gf, 31.15.X-}

\keywords{wavelets, electronic structure calculations, solid harmonics, invariants, multilinear regression}

\maketitle


\section{Introduction}

Many chemical and physical properties of atomistic systems result from the underlying quantum
mechanical behavior of the electrons in the system, which are modeled
through the Sch\"{o}dinger equation with molecular
Hamiltonian. However, 
computing numerically 
the wave function \cite{bartlett:cctqc2007}, the electronic density \cite{hohenberg:dft1964, PhysRev.140.A1133}, and other molecular properties, requires significant computational resources that restrict the size and number of atomistic systems
that can be studied. 
This has led to
substantial
interest recently in developing 
machine learning 
techniques
that can efficiently and accurately 
predict
chemical properties of atomistic
systems 
after being trained on examples where these properties are known.
The scope of these approaches
ranges from estimating potential energy surfaces \cite{MATS:MATS040030207, doi:10.1021/jp055253z,
  Behler:NNPotEnergy2007,
  :/content/aip/journal/jcp/134/7/10.1063/1.3553717,
  bartok:gaussAppPot2010,
  :/content/aip/journal/jcp/139/18/10.1063/1.4828704,
  bartok:repChemEnviron2013, Szlachta:gaussianTungsten2014,
  PhysRevLett.114.096405, :/content/aip/journal/jcp/144/3/10.1063/1.4940026,
  Shapeev2015-MTP, chmiela:MLconservativeFF2017},
to methods that estimate a variety of
properties of molecular configurations spread
across chemical compound space
\cite{rupp:coulombMatrix2012, hansen:quantumChemML2013,
montavon:mlMolProp2013, ramakrishnan:mlTDDFT2015, hansen:BoB2015,
C6CP00415F, schutt:qcDeepTensor2016, gilmer:MPNNQC2017,
schutt:molecuLeNet2017}.

These machine learning algorithms 
are designed to output predictions
that respect the underlying
physical laws of the system. In particular, global chemical and physical properties
of atomistic systems are invariant to rigid movements,
and vary smoothly with respect to the change of single atomic positions. 
In addition, chemical systems have multiscale structures and interactions.
Systematic treatment of these multiscale properties is absent among the machine
learning methods put forward to date.
This trend is, in part,
due to the types of databases currently being studied in the machine
learning context, e.g, neutral
small organic molecules\cite{Ramakrishnan:2014aa} and materials. In these cases,
the potential energy can be separated into atom centered
components which have localized interactions, and a long range component incorporating
electrostatics and dispersion, which are treated separately \cite{bartok:gaussAppPot2010}. Nevertheless, traits such as
energies, vibrational frequencies, energy gaps  and dipole moments are
complex properties of the atomistic state that are dependent upon all
pairwise interactions in the system. 
These interactions
occur at different scales, e.g. ionic and covalent bonds at short range,
van der Waals interactions at the meso-scale,
and long range Coulomb interactions. 
Furthermore, many chemical properties are not
linearized over scales, but are rather the
result of complex nonlinear coupling of small to large scale interactions
in the system.

In order to build a learned approximation that accounts for the multiscale aspect of the quantum interaction, we extended the two-dimensional scattering transform for quantum energy regression \cite{hirn:waveletScatQuantum2016} to a three dimensional spherical scattering transform \cite{eickenberg:3DSolidHarmonicScat2017}.
The molecule is represented by 
several 3D pseudo-electronic density channels, corresponding to core and valence electrons, approximated by Gaussians centered at atomic locations. 
A translation and rotation invariant representation is computed with a
solid harmonic scattering transform, that separate interactions at different scales 
and across scales, with cascaded wavelet transforms. This is illustrated in Figure \ref{fig: ml model}.

Scattering coefficients have the same invariance and stability properties as 
a quantum energy functional: they are invariant to rigid body motion and stable
to deformations of the input density as proved in \cite{mallat:scattering2012}.
This computational approach has some resemblance with quantum chemical theory.
Similarly to the Density Functional introduced by Hohenberg and Kohn
\cite{hohenberg:dft1964}, which computes the ground state energy for an input density, our procedure takes as input a surrogate density $\rho$ and outputs an approximation of the ground-state energy. It involves solid harmonic wavelets 
which are similar to Gaussian Type Orbitals,  
often used as basis function for molecular orbitals in Kohn-Sham DFT \cite{PhysRev.140.A1133}.
However, in contrast to DFT, we do not fit the wavelets parameters to compute a valid electronic density. We rather use them to parametrize the complex non-linear
relation between the molecule geometry and its energy.

Numerical experiments are performed on the QM9 database \cite{ramakrishnan2014quantum}, which contains ground-state energies of thousands of organic molecules computed with DFT using the hybrid B3LYP functional.  A multilinear regression of ground-state energies from scattering invariants
 achieves a mean absolute error of 0.50 kcal/mol. 
This result is close to state of the art machine learning techniques, in particular those computed with deep neural networks.
In contrast, we
use a much simpler multiscale representation whose mathematical properties are relatively well understood, and which can lead to interpretable results with
sparse linear regressions. Using an orthogonal least squares regression, we obtain an error of 1.96kcal/mol with only 400 coefficients
and 10000 samples. Resamplings show that the first few selected coefficients are always the same, and thus seem to consistently capture chemical information of specific, conditionally ordered levels of importance to the task.

The remainder of the paper is organized as
follows. Sections \ref{sec: multiscale invariants} and \ref{sec: scale interaction
  coefficients} describe the wavelet scattering
coefficients. Section \ref{sec: multilinear regression} details multilinear regressions. Numerical results, discussion and interpretation are
presented in Section \ref{sec: numerical results}. 

\section{Multiscale invariants} \label{sec: multiscale invariants}

Let  $x = \{ (z_k, r_k) \in \Z \times \R^3 \}_k$ be an atomistic state, consisting of atoms with nuclear charges $\{ z_k \}_k$, located at positions $\{ r_k \}_k$.
We denote by $f(x)$ a scalar physical property of the state $x$, e.g, its energy, its HOMO-LUMO gap, or the magnitude of its dipole moment.
To accurately approximate $f(x)$ from examples, we must take advantage of known regularity properties.
We summarize some well known properties of $f(x)$:
\begin{enumerate}[leftmargin=*]
\item Invariance to indexation of the atoms in the state $x$;
\item Invariance to rigid motion of the state $x$;
\item Lipschitz continuity with respect to relative variations of distances
between atoms; 
\item Dependence upon multiscale geometric properties of $x$, with nonlinear couplings between scales.
\end{enumerate}
\begin{figure}
\center
\includegraphics[width=.99\linewidth]{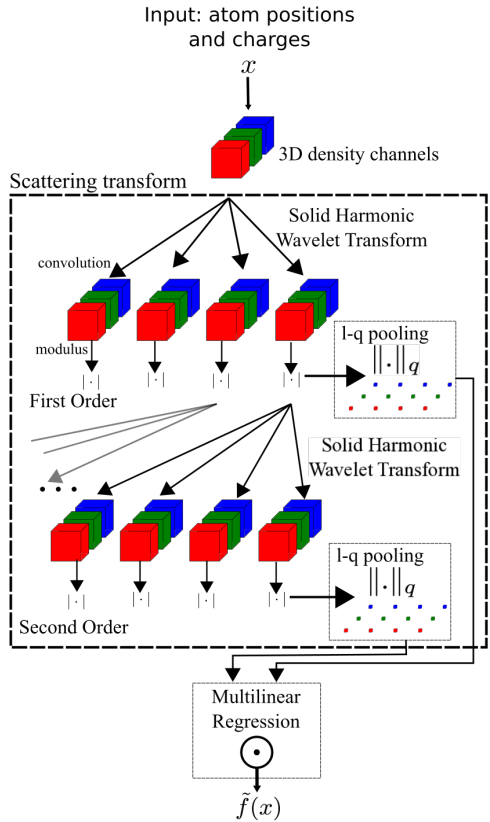}\\
\caption{Scattering regression of a molecular property $f(x)$. The atomistic state $x$ is  represented by several three-dimensional densities.
An invariant scattering transform is applied to each density. 
It computes a solid harmonic wavelet transform of the density, and the modulus of these coefficients are
transformed again by a second solid harmonic wavelet transform. 
Invariant scattering coefficients are spatial $\bf l^q$ norms
of the resulting first and second order modulus densities, for several exponents $q$. 
A multilinear regression computes an estimation $\tilde f(x)$ of
the molecular property from these invariant scattering coefficients. 
}
\label{fig: ml model}
\end{figure}
An approximation $\tilde f(x)$ of the energy functional $f(x)$ is computed
as a multilinear regression over coefficients having the same invariance and stability
properties as $f(x)$. Indeed, invariance properties reduce the
dimensionality of the approximation. 
This is implemented 
with a scattering regression illustrated in Figure \ref{fig: ml model}.
The first step, which is described in Section
\ref{sec: density channels}, maps $x$ to several three-dimensional electronic densities,
denoted here by $\rho_x (u)$. They carry information about the spatial distributions of core and valence electrons, as well as bonds between atoms. These electronic embeddings are invariant to permutation of
atomic indexation. The second step computes an invariant
scattering transform $S \rho_x$ of
each density channel. It separates the variations of $\rho_x$ at different scales, 
and includes scale interaction terms. These coefficients are Lipschitz continuous to deformations of $x$, and hence to relative variations of distances between atoms.
The computation of these coefficients is
described in Sections \ref{sec: solid harmonic wavelets}, \ref{sec: first order scat coeffs} and \ref{sec: scale interaction coefficients}. These coefficients satisfy the  properties listed above. A multilinear regression, described in Section \ref{sec: multilinear regression}, 
computes the estimation $\tilde f(x)$ of the molecular property $f(x)$.  Regression coefficients are optimized from the scattering transforms of
a training set of molecules with their molecular values
 $\{ S \rho_{x_i}, f(x_i)) \}_{i \leq n}$.

\subsection{Three dimensional densities} \label{sec: density channels}

The Hohenberg-Kohn theorem\cite{hohenberg:dft1964}
implies 
that the ground
state energy of an electronic system can be computed as a functional of the electronic density. DFT for molecules relies on this and models the electronic density as the square sum of a linear combination of many Gaussian-type orbitals centered on atom positions.
We represent the state $x$ by several density surrogates, computed as sums of non-interacting densities which depend upon the position and charge of each atom.  
This initial embedding 
ensures that the resulting prediction $\tf (x)$ will be
invariant to permutations of the indexation of the atoms (property 1). 

We compute three non-interacting density channels: one for core electrons, one for valence electrons, and a total density, which is the sum of the core and valence densities. 
These densities are computed as sums of non-interacting single-atom density functions placed at the atom locations.
We use Gaussians to model these single atom densities, scaled to integrate to the amount of charge they represent.
A three dimensional density is computed as:
\begin{equation*}
\rho_x (u) = \sum_k \gamma_k \,g(u - r_k),
\end{equation*}
where $\gamma_k$ is a number of electrons at atom location $r_k$ and $g(u)$ is a three-dimensional Gaussian normalized to have unit integral. 

Core and valence
densities are obtained by setting $\gamma_k$ to be the number of core electrons
or the number of valence electrons of atom $k$. 
The total electronic density is the sum of the core and valence
densities:
\begin{equation*}
\rho_x^{\mathrm{total}} (u) = \rho_x^{\mathrm{core}} (u) +
\rho_x^{\mathrm{valence}} (u).
\end{equation*}
Due to the normalizations of the core and valence densities, the total
density integrates to the total charge of the system: $\int
\rho_x^{\mathrm{total}} (u) = \sum_k z_k$. 

To disentangle bond information from atom type, we create an additional channel to encode atomic bond information explicitly.
This is a surrogate density concentrated along line segments between pairs of bonded atoms, whose integral is equal to the bond order.
Let $B = \{ (i, j) \mid \text{atom } i \text{ and } j \text{ are bonded}\},$ and
$$\rho_x^\textrm{bonds}(u) = C\, \sum_{(i, j)\in B}|r_i - r_j |^{-1}\, 
\gamma_{ij}\,\exp(-d_{ij}^2(u)/2d_0^2)$$
where $d_{ij} (u)$ is the distance between $u$ and the segment
between the atoms $i$ and $j$, $d_0$ is a characteristic width,
$\gamma_{ij}$ is the number of electrons involved in the bond and $C$
is a constant so that 
$\int \rho_x^\textrm{bonds}(u)\,du = \sum_{ij} \gamma_{ij}$.

\begin{figure}
\center
\includegraphics[width=.99\linewidth]
{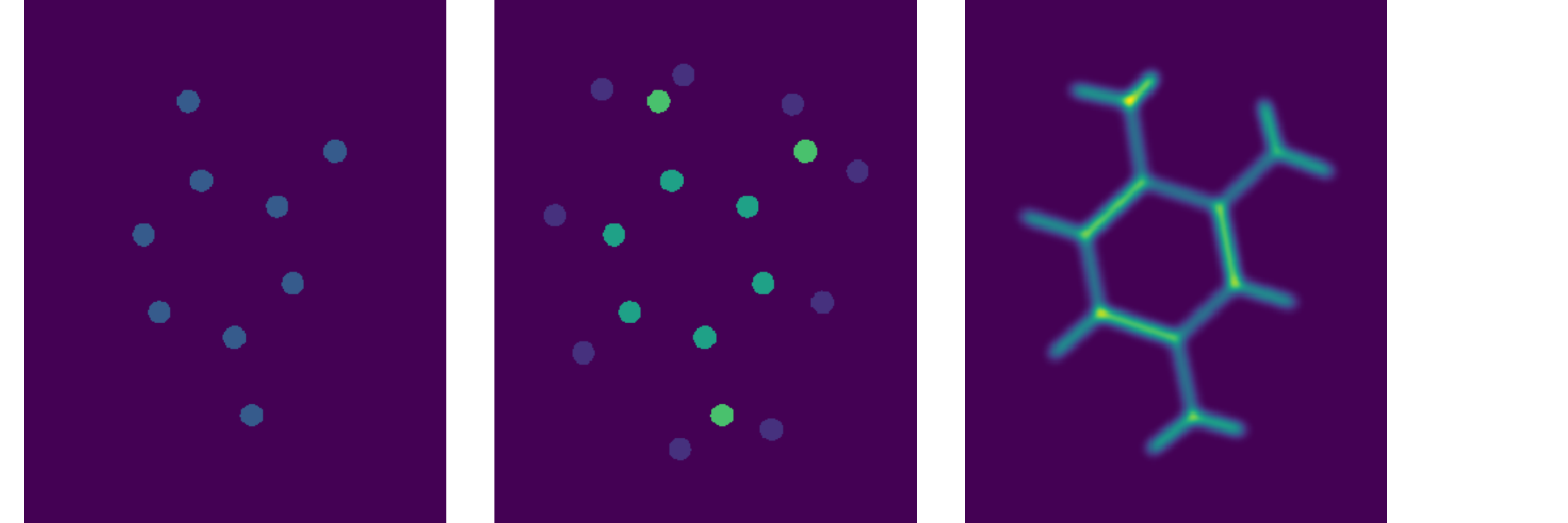}\\
\caption{Slice plots of the core electronic and valence densities thresholded, color-coded by charge count, and bond density of a planar molecule.
The core density does not show Hydrogen atoms.}
\label{fig: density channels}
\end{figure}

All
density channels provide unique geometric and chemical information about the molecule.
Indeed, the position on the two axes of the periodic table, which encodes a large amount of chemical variability, is accounted for in the valence and core channels. This gives direct access to the properties that vary along these axes, such as distance of the outer orbitals to the nucleus and number of available electrons for bonding. The full density channel gives direct access to the actual atom identity, which can help encode its typical purposes within the molecules. The bonds channel can, when this information is available, provide the additional structure indicator of how many valence electrons actually bonded.
Figure \ref{fig: density channels} plots a slice of each of the core, valence and bonds
channels of $\rho_x$ for an organic molecule with planar symmetry.
An alternative, more detailed approach would be to model the densities for small submolecular structures involving more than one atom as described in \cite{gastegger2017wacsf}.

\subsection{Scale separation with solid harmonic wavelets} \label{sec: solid harmonic wavelets}

We separate the different scale components of each density channel by computing a  solid harmonic wavelet transform. 
In the following we write $\rho$ for any density channel
$\rho_x^{\alpha}$ of an arbitrary state $x$.

A wavelet transform integrates the density function $\rho(u)$ against
a family of localized oscillating waveforms, called wavelets, which
have zero mean and are dilated to different scales. The resulting
wavelet coefficients give a complete representation of $\rho$ (and
hence $x$), but reorganize its information content according to the
different scales within the system. 

\begin{figure}
\center
\includegraphics[width=.99\linewidth]{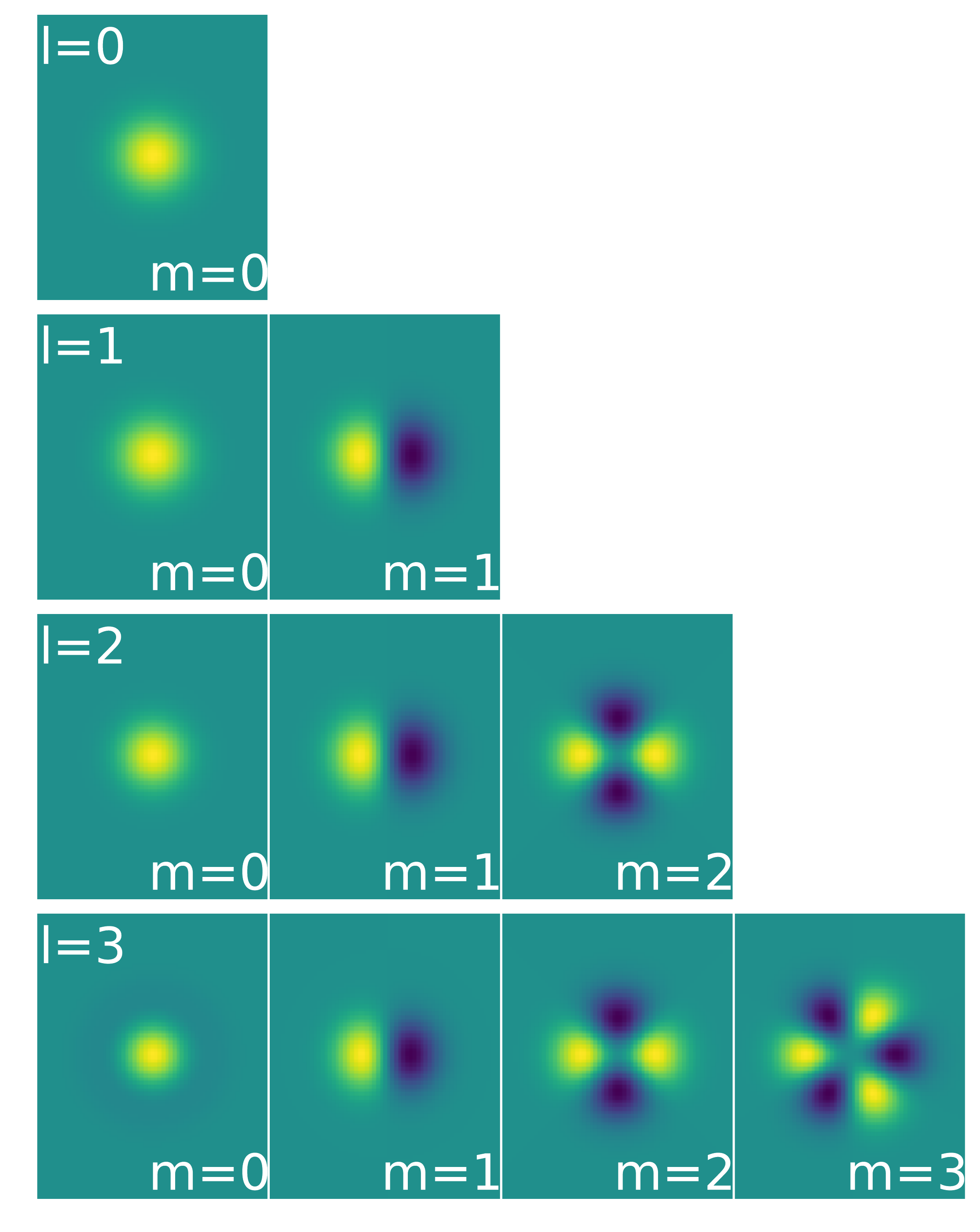}
\caption{Solid harmonic wavelets real parts at fixed scale viewed on an off-center slice at fixed x-value. For a given $\ell$, the global order of oscillation is always the same, but across different values of $m$, the oscillations are traded off between the visible plane and the sliced z-dimension. The planar cuts are also circular harmonics.}
\label{fig: wavelets}
\end{figure}

We use solid harmonic wavelets, first introduced in
\cite{eickenberg:3DSolidHarmonicScat2017} in the context of molecular energy regression, but which have also been used in generic 3D image processing \cite{reisert:3DharmonicFilters2009}. They are defined as
\begin{equation*}
\psi_{\ell}^m (u) = \frac{1}{(\sqrt{2\pi})^3} \,e^{-|u|^2/2}\, |u|^{\ell}\,
Y_{\ell}^m \,\left( \frac{u}{|u|} \right),
\end{equation*}
where $\{ Y_{\ell}^m : \ell \geq 0, -\ell \leq m \leq \ell \}$
are the Laplacian spherical harmonics. Solid harmonic wavelets are solid
harmonics, $|u|^{\ell}
Y_{\ell}^m ({u/|u|})$, multiplied by a Gaussian, which localizes
their support around 0. Decomposing angular frequencies using spherical harmonics will facilitate the computation of rotational invariants, which are described in Section \ref{sec: first order scat coeffs}. Dilations of $\psi_{\ell}^m$ at the scale $2^j$ are given by:
\begin{equation*}
\psi_{j,\ell}^m (u) = 2^{-3j} \psi_{\ell}^m (2^{-j} u).
\end{equation*}
Figure \ref{fig: wavelets} plots slices of their response in $\R^3$.

Solid harmonic wavelets are of the same functional form as Gaussian-type
orbitals (GTOs) \cite{Boys542}, which are often used in computational quantum chemistry, 
e.g. as basis sets for the Kohn-Sham orbitals in DFT. In these approaches one electronic density may be
made up of a linear combination of hundreds of atom-centered GTOs, in
which several GTOs are used to model each electron orbital of each
atom and the corresponding electronic interactions that make up the
bonds in the system.

Here, solid harmonic wavelets are used as convolution filters.  A
solid harmonic wavelet transform integrates the density $\rho(v)$ against each
solid harmonic wavelet $\psi_{j,\ell}^m(v)$ centered at each spatial location
$u \in \R^3$:
\begin{equation*}
W \rho = \Big\{ \rho \ast \psi_{j,\ell}^m (u)  =
\int_{{\mathbb R}^3} \rho(v) \,\psi_{j,\ell}^m (u-v)\, \textrm{d}v \Big\}_{j,\ell,m}
\end{equation*}
where the scales go from $j = 0, \ldots, J-1$, and the angular
frequency bands go from $\ell = 0, \ldots, L-1$, for prescribed values of
$J$ and $L$. The wavelet coefficients $W \rho$ separate the geometry
of $\rho$ into different scales $j$ and angular frequency bands
$\ell$. The additional frequency variable $m$ further subdivides the
angular frequency information, while the spatial variable $u$ is
necessary due to the localization of the waveform, and encodes the
response of $\rho$ against $\psi_{j,\ell}^m$ at this location.

For an electronic density $\rho_x (u) = \sum_k \gamma_k \,g(u-r_k)$,
\begin{equation*}
\rho_x \ast \psi_{j,\ell}^m (u) = \sum_k \gamma_k \,g \ast
\psi_{j,\ell}^m (u - r_k).
\end{equation*}
Since $g \ast \psi_{j, \ell}^m$ is another (wider) Gaussian-type
orbital, we may view these coefficients as emitting the same GTO from
each atom, scaled by the electronic charge $\gamma_k$.
These coefficients encode interferences of
the solid harmonic waveforms at different scales. 
For small scales,
these interferences resemble those found in the computation of
electronic orbitals in computational chemistry software. Solid harmonic
wavelet coefficients resulting from larger scale waveforms aggregate
subgroups of atoms within the system, and encode macroscopic geometric
patterns.
While these coefficients do not have direct analogues in classical density estimation procedures,
they are required for deriving invariant coefficients that can account for long range interactions within the system.

\subsection{Solid harmonic wavelet invariants} \label{sec: first order scat coeffs}

Although the solid harmonic wavelet
transforms
$\{\rho\ast\psi_{j,l}^m(u)\}$
form a complete representation of
$\rho$, they are not suitable 
as input to a regression for
global chemical
properties. Indeed, a translation of $\rho$ (obtained by shifting the spatial input variable $u$, $\tau_v\rho(u) = \rho(u - v)$) results in a translation
of the wavelet coefficients. 
Rotations of $\rho$ similarly rotate the
variable $u$, but also redistribute the wavelet coefficients across
the $m$ angular frequencies within each angular frequency band
$\ell$. 
Using these coefficients 
for
regression would require the regression to learn all invariance properties that 
are already known,
entailing a very high sample complexity.
We address this point by computing coefficients with the same invariances as the target model. In order to obtain invariant coefficients to rotations of the state $x$, we first aggregate the energy of the solid harmonic wavelet coefficients across $m$:
\begin{equation*}
U[j,\ell] \rho (u) = \left( \sum_{m = -\ell}^{\ell} | \rho \ast
  \psi_{j,\ell}^m (u)|^2 \right)^{1/2}.
\end{equation*}
The value $U[j,\ell] \rho$ is computed with a Euclidean norm that we shall call a 
\textit{modulus operator}. It eliminates the rotational phase subspace information from the wavelet coefficients, and
one can prove that the resulting coefficients are covariant to both translations and
rotations (meaning that a translation or rotation of $\rho$ results in the
same translation or rotation of $U[j,\ell] \rho$). See Figure \ref{fig:
  wavelet modulus coeffs} for plots of $U[j,\ell] \rho_x$.

\begin{figure}
\center
\includegraphics[width=.99\linewidth]{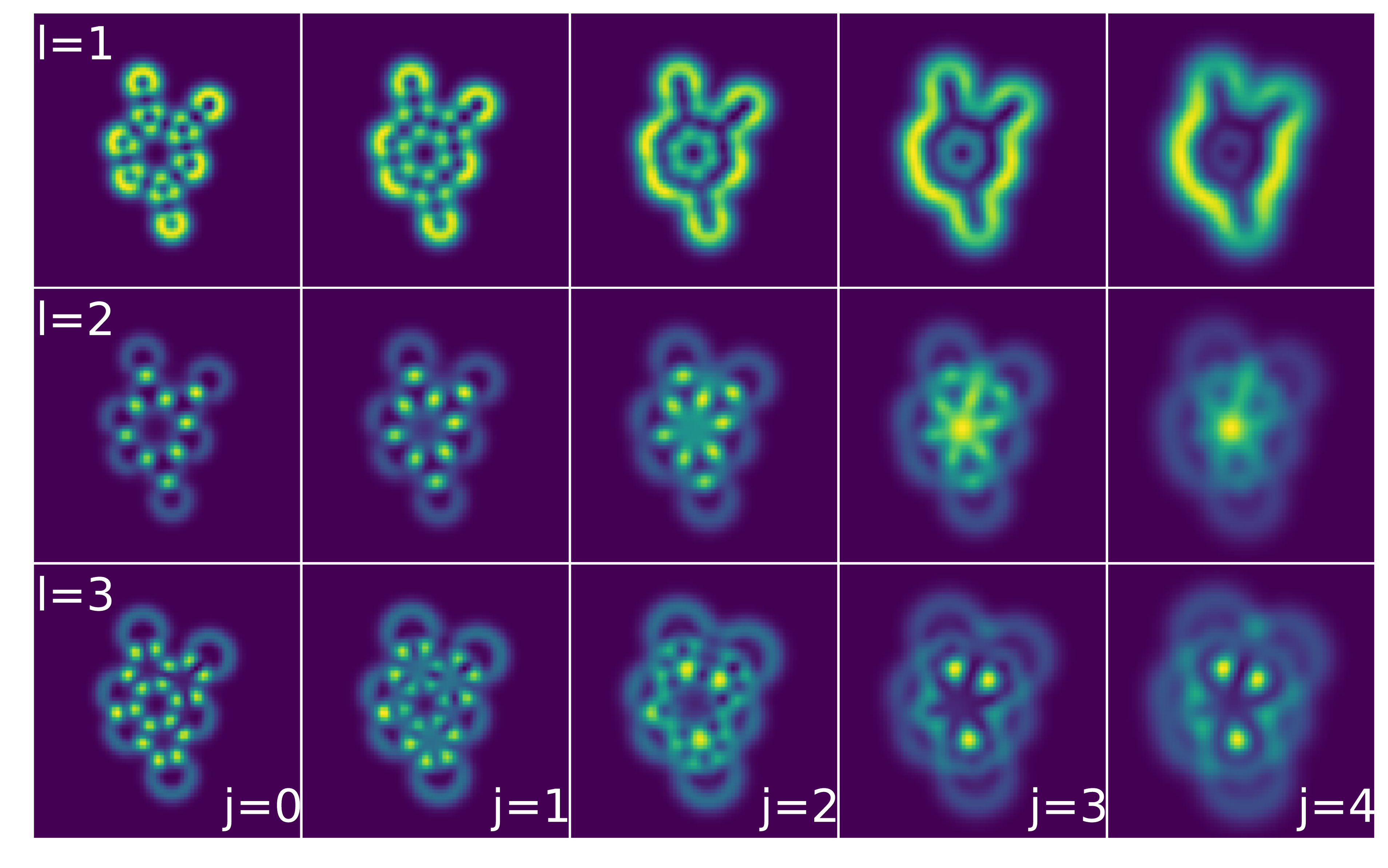}
\caption{Solid harmonic wavelet modulus coefficients $U[j,\ell]
  \rho^\textrm{valence}_x$ of a planar molecule from the QM9 dataset, for $\ell=1, 2, 3$ oscillations and five scales. Depending on angular frequency and scale, different interference patterns arise, which capture complementary aspects of the molecule}
\label{fig: wavelet modulus coeffs}
\end{figure}

It follows from this covariance property that
rigid motion invariant coefficients may be obtained by integrating
over $u$:
\begin{align*}
S \rho [j,\ell,q] &= \int_{\R^3} | U[j,\ell] \rho (u)|^q \, \textrm{d}u, \\
&= \int_{\R^3} \left( \sum_{m = -\ell}^{\ell} | \rho \ast
  \psi_{j,\ell}^m (u)|^2 \right)^{q/2} \, \textrm{d}u.
\end{align*}
We refer to $S \rho [j,\ell,q]$ as first order solid harmonic
wavelet scattering coefficients. Each value $S \rho [j,\ell,q]$
collapses $U[j,\ell] \rho$ (illustrated by a single box in Figure
\ref{fig: wavelet modulus coeffs}) into a single coefficient. They
give invariant descriptions of the state $x$ that are stable to
deformations of the atomic positions. Furthermore, they separate the
scales of the system by aggregating geometric patterns related to the
arrangement of the atoms and bonds at each
scale. 

The power $q=1$ scales linearly with the number of electronic particles represented by the density $\rho$, while $q=2$ encodes pairwise interactions, which are related to electrostatic Coulomb interactions in \cite{hirn:waveletScatQuantum2016}. 
Specifically, the wavelet modulus at $\ell=1$, $U[j, 1]\rho$, computes the gradient magnitude of the density at  scale $j$. When integrated over space with $q=1$, this results in the total variation of the scaled density, a quantity well-known in functional analysis. It measures the lengths of density level sets with emphasis on inflection lines. At a sufficiently large scale, it measures the perimeter of the molecule (see the first line of figure \ref{fig: wavelet modulus coeffs}).  
Powers smaller than $1$ provide sparsity information: A small exponent $q$ gives more importance to small non-zero values, essentially counting them. Large exponents are sensitive to peaks of large amplitudes.

First order solid harmonic scattering coefficients initially encode the state $x$ in a family of Gaussian densities $\rho_x$, and use spherical harmonics to obtain rotational invariants; in this regard they are similar to the smooth overlap of atomic positions (SOAP) algorithm introduced in \cite{bartok:repChemEnviron2013}, and subsequently used in \cite{Szlachta:gaussianTungsten2014,C6CP00415F}. SOAP representations, and resulting similarity measures, however, are computed on localized atom -centered molecular fragments at a fixed scale, which are then combined to form global similarity measures between different molecules \cite{C6CP00415F}. First order scattering coefficients $S \rho_x [j,\ell,q]$ are global integrated descriptions of the state $x$, which are computed at each scale $2^j$.

\section{Scale interaction coefficients} \label{sec: scale interaction
  coefficients}

First order scattering coefficients separate the scales of the system
$x$. They encode both micro- and macroscopic invariant geometric
descriptions of $x$, corresponding to short- and long-range
interactions. However, the chemical and physical properties of complex
quantum systems with many particles are the result not only of interactions at
each scale, but the nonlinear mixing of scales (as in London dispersion
forces \cite{london:LDF1937}). We thus complement first order scattering coefficients with
second order scattering coefficients that combine information across scales within the system.
If a certain sub-molecule property becomes visible to a solid harmonic scattering modulus at a given scale, then interactions between these sub-molecule units can be captured by analyzing this modulus at a larger scale.

Solid harmonic wavelet modulus coefficients $U[j,\ell]\rho$ are decomposed with a second wavelet modulus transform, with wavelets $\psi_{j',\ell}^m$ that are larger than the first wavelet:
\begin{equation*}
U[j,j',\ell]\rho (u) = \left( \sum_{m=1}^{\ell} | U[j,\ell]\rho \ast \psi_{j',\ell}^{m} (u)|^2 \right)^{1/2}, \quad j < j'. \\
\end{equation*}
The resulting coefficients mix relatively small scale geometries of $\rho$ across the larger scale $2^{j'}$, in a distinctly different fashion than, for example, computing the product $S\rho[j,\ell,q] S\rho[j',\ell,q]$.
Indeed, wavelet modulus coefficients $U[j,\ell]\rho$ are covariant to rotations and translations, and thus maintain oriented geometric information within the state $x$ at the scale $2^j$.
This information is analogous to localized multipole moments, such as local dipoles that arise due to the polarizability of atoms or sub-molecules within the state $x$.
These weak localized interactions within the state $x$, in aggregate, non-trivially impact the large scale structure of the system.
The second wavelet $\psi_{j',\ell}^m$ sets the scale at which this structural impact is measured, and the localized contributions at the smaller scale are aggregated via integration:
\begin{equation*}
S\rho [j,j',\ell,q] = \int_{\R^3} |U[j,j',\ell]\rho(u)|^q \, du.
\end{equation*}

The resulting coefficients $S\rho [j,j',\ell,q]$ are second order solid harmonic wavelet scattering coefficients, which are depicted in Figure \ref{fig: ml model}. These coefficients are invariant to rigid motion of $\rho$, stable to deformations of the state $x$, and couple two scales within the atomistic system. We conjecture that the resulting terms are qualitatively similar to first-order approximations of van der Waals interactions, which are given by $C \alpha_1 \alpha_2 / R^6$, where $\alpha_k$, $k=1,2$, are the polarizabilities of two local substructures within $x$ that are well separated by a relatively large distance $R$. A second order scattering coefficient can potentially capture such a term if the substructures are of size proportional to $2^j$ and $R \sim 2^{j'}$.
 
These van der Waals terms, in addition to electrostatic terms and localized energy terms, are utilized in expansions of the total energy of $x$ via perturbation theory. Since we do not have a direct bijective correspondence with such terms, we instead learn and compute such energy expansions via multilinear regressions, which are described in the next section.

While the described forces play a subordinate role in the systems we study in Section \ref{sec: numerical results}, we find empirically that the inclusion of the described scale interaction terms significantly increases the predictive power of our analysis method. Furthermore, we expect that for systems with significant van der Waals interactions that second order scattering coefficients will be necessary to achieve reasonable regression errors. The mathematical and empirical study of this hypothesis is the subject of ongoing and future work.

\section{Multilinear regression} \label{sec: multilinear regression}
  
Physical properties of atomistic states are regressed with multilinear
combinations of scattering coefficients $S \rho_x$. A
multilinear regression of order $r$ is defined by:
\begin{equation*}
\tf_r(x) = b + \sum_i (\nu_i \prod_{k=1}^r (\langle S\rho_x,
w_i^{(k)}\rangle + c_i^{(k)})).
\end{equation*}
For $r = 1$, it gives a linear regression that we write by expanding the inner product:
\begin{equation} \label{eqn: linear model}
\tf (x) = b + \sum_{\substack{j,\ell,q}} w_{j}^{\ell,q} S\rho_x [j,\ell,q] + \sum_{\substack{j'> j}} w_{j,j'}^{\ell,q} S\rho_x [j,j',\ell,q] .
\end{equation}
where different scales $j$ and $j'$, oscillation numbers $\ell$ and integration exponents $q$ are separated in different coefficients.
For $r=2$ this form
introduces a non-linearity similar to those found in factored gated autoencoders \cite{memisevic2011gradient}.
We optimize the parameters of the multilinear regression by minimizing the quadratic loss over the training data,
\begin{equation*}
\sum_{i=1}^n \left( f(x_i)-\tf_r(x_i) \right)^2,
\end{equation*}
using the Adam algorithm for stochastic gradient descent \cite{kingma2014adam}.

If $f (x)$ is the total energy of the state $x$, and $x$ is decomposable into a number of simpler subsystems, then perturbation theory analysis expands the energy $f(x)$ into an infinite series of higher order energy terms:
\begin{equation*}
f(x) = f_0(x) + f_1(x) + f_2(x) + \cdots,
\end{equation*}
in which $f_0$ captures the energies of the isolated subsystems, $f_1$ captures their electrostatic interactions, and $f_2$ captures induction and dispersion energies, which result from van der Waals interactions. The linear regression \eqref{eqn: linear model} similarly expands the total energy into successively higher order energy terms, but which are defined by wavelet scattering coefficients. As described in the previous sections, first order scattering coefficients for $q=2$ likewise encode pairwise Coulombic interactions, while second order scattering coefficients couple scales within the system, analogous to first order approximations of induction and dispersion energies.

\section{Numerical results} \label{sec: numerical results}

We train the regression algorithm to predict energies and other properties of organic molecules with first- and second-order
solid harmonic wavelet scattering coefficients. Detailed numerical estimates and comparisons with the state
of the art are presented in Section \ref{sec: numerical regression
  errors}. In Section \ref{sec: interpretation} we interpret
scattering embeddings of organic molecules by analyzing sparse
linear regressions.

Software to reproduce the results is available at \texttt{http://www.di.ens.fr/data/software}
\subsection{Numerical regression errors} 
\label{sec: numerical
  regression errors}
We use our algorithm to approximate molecular energies, i.e. atomization energy at $0K$  $U_0$, atomization energy at room temperature $U$, enthalpy of atomization at room temperature $H$ and atomization free energy at room temperature $G$.
We use the values of the QM9 dataset \cite{ramakrishnan:mlTDDFT2015}, which contains
133,885 different small organic molecules with maximally 29 atoms of the elements C, H, O, N, and F, and maximally 9 heavy (non-Hydrogen) atoms. Their molecular configurations and properties were estimated using B3LYP DFT.

First- and second-order scattering invariants were computed for core, valence, full, and bonds channels, with $J=4$, and $\ell\in\{1, 2, 3\}$. The moduli were raised to the powers $q\in\{1/2, 1, 2, 3, 4\}$.

Multilinear regressions were trained and evaluated on five random splits of the dataset, for $r=1$, i.e. linear regression, and $r=3$, trilinear regression. The evaluation criterion for the fits was mean absolute error, which is the most prevalent error measure in the literature. 
We use 5-fold cross-validation, separating the data in 107,108 training datapoints and 26,777 test datapoints per fit.

For the atomization energy at room temperature $U$, trilinear regression achieves close to state of the art error rates on core, valence and full densities (0.56 kcal/mol). When incorporating bond information via the bonds channel, the result improves to  0.51 kcal/mol. This is expected, since the bonds channel contains relevant information pertaining to known bond structure, which are not immediately deducible from the number of valence electrons of the constituent atoms.

Notably, the error of a simple linear regression is much lower than that of Coulomb matrices fit with kernel ridge regression at full sample complexity.

\begin{table}[]
\centering
\caption{Prediction errors for molecular energies of the QM9 dataset in kcal/mol.
From right to left: the scattering with linear regression, the scattering with trilinear regression, Neural Message Passing and Coulomb Matrices. 
}
\label{table: properties regression}
\begin{tabular}{|c|c|c|c|c|c|c|}
\hline
          & L-Scat & T-Scat   &  NMP  & CM    & SchNet \\ \hline
$U_0$     & 1.89   & 0.50     &  0.45 & 2.95  & 0.31  \\ \hline
$U$       & 2.4    & 0.51     &  0.45 & 2.99  & {}    \\ \hline
$H$       & 1.9    & 0.51     &  0.39 & 2.99  & {}    \\ \hline
$G$       & 1.87   & 0.51     &  0.44 & 2.97  & {}    \\ \hline
\hline
$\mu$      &     0.63  & 0.34 &  0.030  & 0.45    & {}      \\ \hline
$\alpha$   &      0.52   &    0.16  &  0.092 & 0.43     & {}      \\ \hline
$\epsilon_{\text{HOMO}}$    &   4.08  &  1.97  & 0.99  & 3.06    & {}     \\ \hline
$\epsilon_{\text{LUMO}}$    &    5.39   & 1.76  & 0.87  & 4.22   & {}    \\ \hline
$\epsilon_{\text{gap}}$  &       7    &   2.73   & 1.60 & 5.28   & {}     \\ \hline
$\langle R^2\rangle$  &  6.67       &  0.41      &  0.18 & 3.39   & {}      \\ \hline
zpve        &  0.004    &  0.002    &   0.0015 & 0.0048   & {}      \\ \hline
$C_v$       &  0.10    &   0.049         & 0.04 & 0.12   & {}     \\ \hline
\end{tabular}
\end{table}

The QM9 dataset contains eight other molecular properties that we are able to predict. However, our technique 
has been primarily optimized to regress the energy properties which are at the top of Table \ref{table: properties regression}. 
The bottom part provides prediction errors for the dipole 
moment $\mu$ (Debye), the static polarizability $\alpha$ ($a_0^3$, Bohr radius cubed), the $\epsilon_{\text{HOMO}}$  energy (kcal/mol), 
the $\epsilon_{\text{LUMO}}$  energy (kcal/mol), the $\epsilon_{\text{gap}}$ gap energy (kcal/mol), the electronic spatial extent $\langle R^2\rangle$  
($a_0^2$), the zero point vibrational energy (eV), and the heat capacity at room temperature $C_v$ (cal/mol/K).
The results show
good
prediction performance on all properties of the molecules. Some error values are close to state of the art, and the majority no further than a factor 2 larger in error. 

Using linear regression we checked the impact of omission of one or several of the input channels. While the valence atomic channel accounts for the strongest change in variance explained, both core and full density channels are required to achieve satisfactory results. The additional accuracy provided by the bonds channel is smaller, but significant.
Notably, comparing the linear regression error incorporating all four channels (1.89kcal/mol) to the four linear regression models in which each channel is left out in turn, we obtain the following error reductions: Removing the full density incurs 0.64 kcal/mol, removing the core density incurs 0.51kcal/mol, removing the valence channel incurs 0.73kcal/mol, and removing the bonds channel results in an error increase of 0.16kcal/mol from the joint prediction error of 1.89kcal/mol. Removing any of the channels results in a non-negligible error increase.
Future directions include the exploration of molecular dynamics via force derivatives of the energy functional. In this setting, the bonds channel will be omitted, since in that setting bonds are subject to temporal variation and may be broken or created.

\subsection{Scattering model interpretation} \label{sec: interpretation}
\begin{figure}
\center
\includegraphics[width=.49\linewidth]{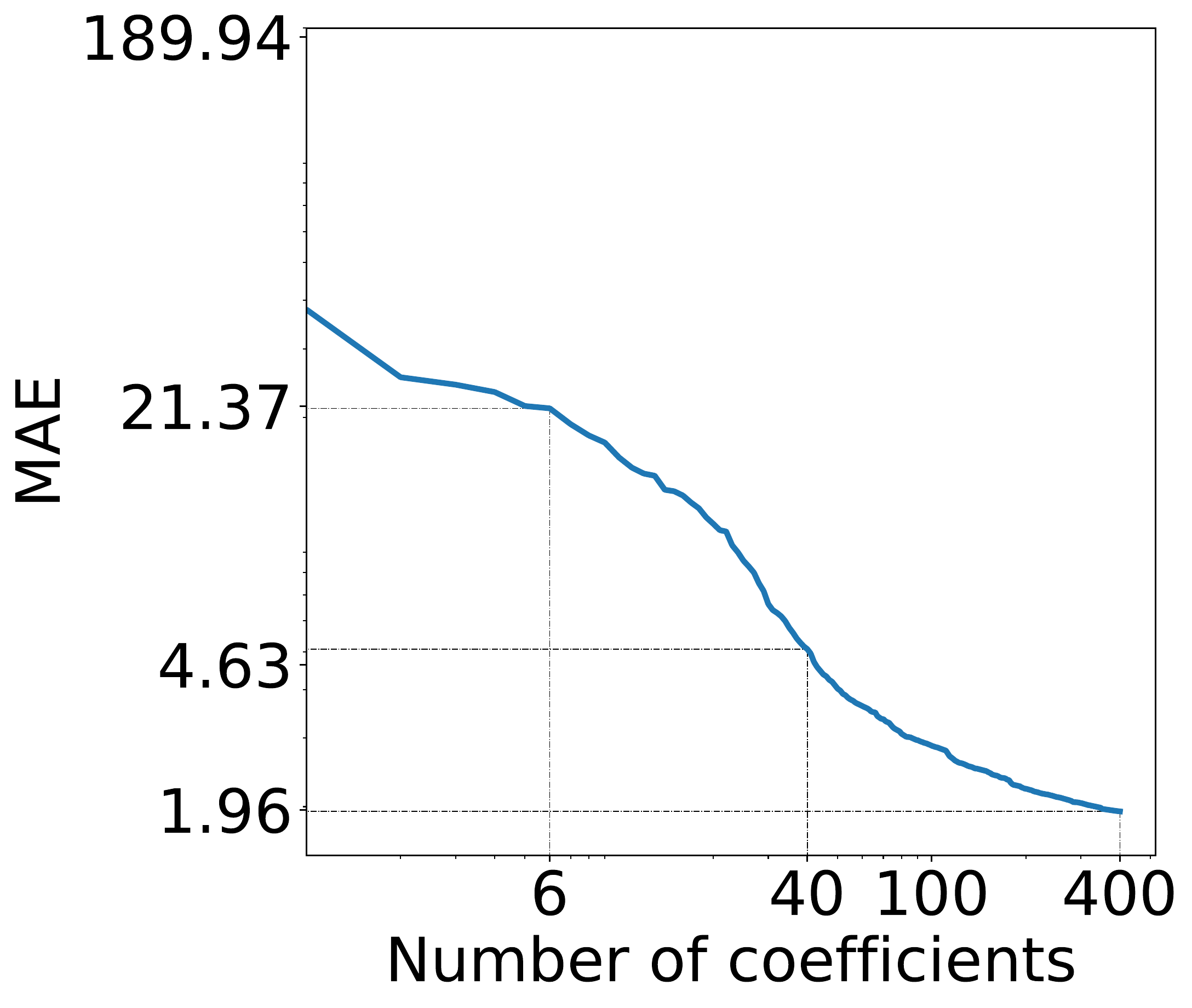}
\includegraphics[width=.49\linewidth]
{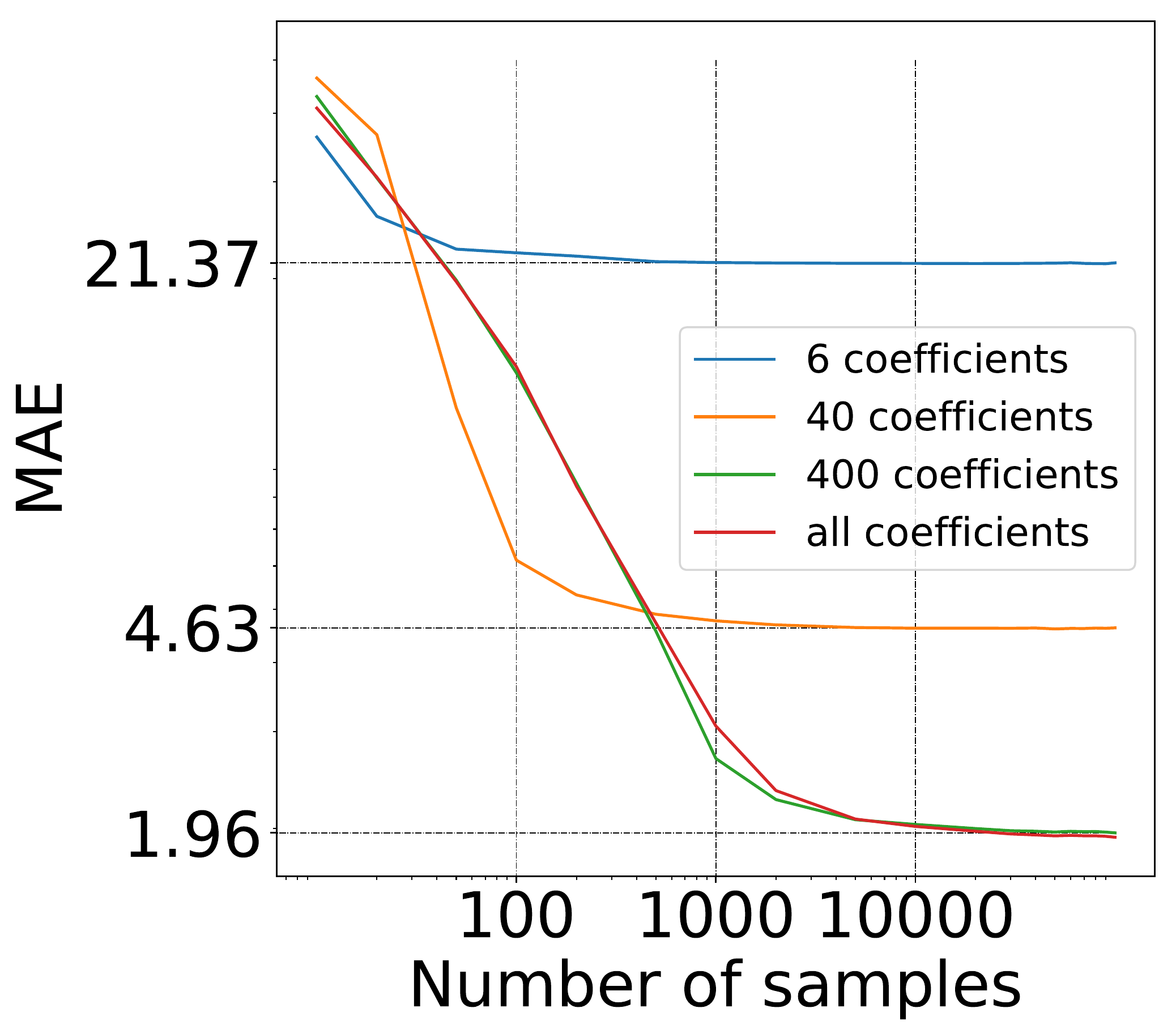}
\caption{\textbf{Left:} Mean absolute error of Orthogonal Least Squares regression for property $U$ on a held-out test set as a function of sparsity level. For every property, the first six coefficients selected are always the same across random splits.
Error decreases rapidly for sparse regressions up to 40 coefficients.
Sparse regression with 400 coefficients is close to or better than a full linear regression. \textbf{Right:} MAE prediction error for three fixed sparse OLS regressions (6, 40, and 400 coefficients) and a full linear regression as a function of number of samples. Different numbers of coefficients lead to different bias-variance tradeoffs. Smaller numbers of coefficients require few samples to train, but cannot improve given more samples. Larger numbers of coefficients require more samples, but the error decreases further.}\label{fig:mae error}
\end{figure}

Linear regressions from scattering invariants have good prediction properties. We use this
fact to compute
small-size interpretable models, with sparse regressions. These sparse regressions are computed with an Orthogonal Least Square (OLS) pursuit, which is a greedy algorithm that decorrelates regressors. 
At every iteration this algorithm\cite{hirn:waveletScatQuantum2016} selects the scattering coefficient that predicts the target best, and then orthogonalizes the system with respect to that predictor. It gives us a list of coefficients ordered by conditional importance.
We apply this procedure to different splits of the dataset, which can 
yield different coefficient lists if selection is unstable (e.g. when two features add similar amounts of predictive power, such that the randomness of data splitting can make either one be selected first).
Figure \ref{fig:mae error} (left) shows the average error for atomization energy $U$ w.r.t the number of features selected by OLS.

We can see clearly that the more coefficients are included in a regression, the further the estimation error decreases.
In particular, we see a strong initial drop in error with each new coefficient and with only 40 coefficients, we are already within range of the Coulomb Matrix reference error.
The error continues to monotonically decrease to 1.96	kcal/mol at 400 coefficients.

We also analyze how many samples are needed for the training: with 6 coefficients, only 100 samples are needed to train almost fully, but the error cannot go under 21 kcal/mol; with 40 coefficients, the error asymptote of 4.63 kcal/mol is reached at roughly 1000 samples. With 400 coefficients, the performance is similar to the full linear regression, and the error asymptote of 1.96 kcal/mol is reached at around 10,000 samples.

Given this knowledge we can conclude that our current linear estimators can be trained on very small subsets of QM9. Indeed, our sparse linear models require considerably less data
when compared to scattering trilinear regressions and current state of the art -  message passing neural network algorithms and SchNet. Furthermore, while kernel methods with appropriate kernels may be able to decrease the error to 0 given a large or infinite amount of data, their representation sizes increase with the number of samples. Using fixed-size and especially linear regressions, as we do, we are able to detect their capacity limit and use the relevant prediction depending on the case.

Another direction of extrapolation relevant to predictions evaluation is molecule size. Ideally, a machine learning technique should be able to predict properties of molecules of a different size from those on which it was trained. An estimator with this property could be useful for computing chemical properties of larger-scale structures. Defining molecule size by the number of heavy atoms it contains, we created training sets of molecules with up to 7, up to 8 and up to 9 heavy atoms. We evaluate predictive accuracy of the linear regression on test sets likewise composed of up to 7, 8, 9 heavy atom molecules. Learning on up to 7 or 8 heavy atoms incurs an error of around 4kcal/mol on molecules with 9 heavy atoms and 2kcal/mol on the smaller ones. Learning with up to 9 heavy atoms gives the usual 2kcal/mol error. It thus seems as though scattering coefficients permit scaling to larger molecule sizes while incurring a constant error. For larger datasets to come, this test will permit clearer conclusions. We believe that such molecular extension properties should be included in the evaluation procedure of machine learning algorithms. 

As stated above, we select the coefficients with OLS using different splits of the dataset.
Notably, the first 6 selected coefficients are always the same, independently of the split of the dataset.
The stability in selection of the same first 6 coefficients invites closer examination.
We apply the same OLS selection of 6 coefficients for the internal energy $U_{int}$, which is equal to the atomization energy $U$ plus the internal energy of all the atoms that form the molecule.
The 6 selected coefficients are shown in table \ref{table: first six coefs}.

\begin{table}[h]

\begin{tabular}{|c|c|c|c|c|c|}\hline
$U$ & density & $j$ & $j'$ & $\ell$ & q\\\hline
1 & bonds & 2 & {} & 1 & 1/2\\\hline
2 & bonds & 0 & 1 & 1 & 1/2\\\hline
3 & bonds & 1 & {} & 2 & 1/2\\\hline
4 & core & 0 & 2 & 3 & 4\\\hline
5 & valence & 3 & {} & 3 & 3\\\hline
6 & full & 0 & {} & 2 & 1/2\\\hline
\end{tabular}
\begin{tabular}{|c|c|c|c|c|c|}\hline
$U_{int}$ & density & $j$ & $j'$ & $\ell$ & q\\\hline
1 & valence & 0 & {} & 1 & 2\\\hline
2 & core & 3 & 4 & 2 & 2\\\hline
3 & full & 3 & 4 & 2 & 2\\\hline
4 & full & 1 & {} & 1 & 3\\\hline
5 & full & 0 & 1 & 1 & 2\\\hline
6 & valence & 0 & 1 & 1 & 4\\\hline
\end{tabular}
\caption{Specification of the first six coefficients used to regress the internal energy $U_{int}$ and the atomization energy $U$. These six coefficients are stable across 16 resamplings.
Note the systematic selection of bonds channels for the estimation of atomization energy.
}\label{table: first six coefs}
\end{table}

Interestingly, we see that the coefficients computed from the bonds density are highly relevant to the atomization energies, but accurate estimation requires information about the nature of the atoms that are contained in the coefficients computed from the core and full densities.
On the other hand, coefficients computed from the full, core and valence density are highly relevant to the internal energy $U_{int}$ of the molecule,
which contains the individual free atom energies as well as the bond energies.
We also see the importance of the second order coefficients (coefficients for which $j' \neq \emptyset$) that account for multiscale interactions.

\section{Conclusion}
\label{sec: conclusion}
We have presented an algorithm that introduces a representation of molecules that is 
invariant to three dimensional rotations and translations.
The algorithm relies on a multiscale representation of molecular densities based on 
solid harmonic wavelets. Solid harmonic wavelets are also used as Gaussian-type orbitals in molecular DFT, establishing an interesting qualitative link between the methods. Our method does not optimize the parameters of the solid harmonics, but rather uses them to analyze surrogate densities created to resemble the molecule electronic density.
The representation is general enough to be applied to any three dimensional density dataset 
and will extend to molecules with a higher number of atoms than the ones presented in the 
training set.
We have shown that the representation can achieve near state of the art predictive 
performance using non linear regression.

Notably, compromising in predictive performance we can use sparse linear regressions to add interpretability to our results. We show that specific coefficients are reliably selected by an orthogonal least squares method across random data splits. The selected coefficients contain bond information when atomization energy is fitted, whose primary fluctuations derive from bonding, and contain atomic information when full ground state energy is fitted, because free atom contributions then dominate the term.
We thus have a precise handle on the interpretation of the coefficients.

\begin{acknowledgements}
M.E., G.E. and S.M. are supported by ERC grant InvariantClass 320959;
M.H. is supported by the Alfred P. Sloan Fellowship (grant FG-2016-6607), the DARPA YFA (grant D16AP00117), and NSF grant 1620216.
\end{acknowledgements}

\end{document}